\documentclass[11pt]{article}%
\usepackage[slantedGreek]{mathpazo}
\usepackage{amsmath}%
\usepackage{amsfonts}%
\usepackage{amssymb}%
\usepackage{graphicx}
\usepackage{subfigure}
\usepackage{multirow}
\usepackage{booktabs}
\usepackage{setspace}
\usepackage{indentfirst}
\usepackage{hyperref}
\hypersetup{ colorlinks=true,       
    linkcolor=red,          
    citecolor=blue,        
    filecolor=magenta,      
    urlcolor=cyan           
    }

\usepackage{natbib}
\newcommand{\iid}{\stackrel{\mathrm{iid}}{\sim}} 
\newcommand{\argmax}{\arg\!\max} 
\newcommand{\argmin}{\arg\!\min} 

\newcommand{\bra}[1]{\left[#1\right]} 

\newcommand{\mat}[1]{\begin{matrix}#1\end{matrix}} 
\newcommand{\bmat}[1]{\bra{\mat{#1}}} 

\begin{document}

\title{Regularizing Bayesian Predictive Regressions \thanks{We appreciate helpful comments from conference participants at 26th Annual Meeting of the Midwest Econometrics Group, Washington University in St. Louis, 2017 Vienna-Copenhagen Conference on Financial Econometrics, 2017 NBER-NSF Seminar on Bayesian Inference in Econometrics and Statistics, and 2017 NBER-NSF Time Series Conference. We thank Jianeng Xu for his excellent research assistance.}}

\author{Guanhao Feng\thanks{%
Address: 83 Tat Chee Avenue, Kowloon Tong, Hong Kong. E-mail address: 
\texttt{gavin.feng@cityu.edu.hk}.} \\
\textit{College of Business}\\
\textit{City University of Hong Kong}
\and Nicholas G. Polson\thanks{%
Address: 5807 S Woodlawn Avenue, Chicago, IL 60637, USA. E-mail address: 
\texttt{ngp@chicagobooth.edu}.} \\
\textit{Booth School of Business}\\
\textit{University of Chicago}}

\date{This Version: September 12, 2017}

\maketitle
\begin{abstract}
\noindent Regularizing Bayesian predictive regressions provides a framework for prior sensitivity analysis via the regularization path. We jointly regularize both expectations and variance-covariance matrices using a pair of shrinkage priors. Our methodology applies directly to vector autoregressions (VAR) and seemingly unrelated regressions (SUR). By exploiting a duality between penalties and priors, we reinterpret two classic macro-finance studies: equity premium predictability and macro forecastability of bond risk premia. We find that there exist plausible prior specifications for predictability for excess S\&P 500 returns using predictors book-to-market ratios, CAY (consumption, wealth, income ratio), and T-bill rates. We evaluate our forecasts using a market-timing strategy and show how ours outperforms buy-and-hold. We also predict multiple bond excess returns involving a high-dimensional set of macroeconomic fundamentals with a regularized SUR model. We find the predictions from latent factor models such as PCA to be sensitive to prior specifications. Finally, we conclude with directions for future research.
\end{abstract}

\begin{flushleft}
\end{flushleft}

\newpage
\section{Introduction}
The Bayesian paradigm in finance and economics requires prior distribution motivated by economic theory, whereas regularization requires a penalty to trade off by optimizing out-of-sample predictive performance. A duality exists between Bayesian methods and statistical regularization which leads to a framework for prior sensitivity analysis. To illustrate our method, we examine the predictability of the equity premium and bond risk premia predictability using macro factors. 

Sensitivity analysis from a Bayesian perspective is typically computationally intensive simulation. However, in economics and finance, performing prior sensitivity analysis across a wide range of prior hyper-parameters at a low computational cost is essential. A significant contribution of our paper is the use of a fast and scalable convex optimization algorithm to perform prior sensitivity analysis. The data-driven method for choosing the tuning parameter, in particular, leads to an alternative interpretation of prior hyper-parameters. For example, we provide a fast sparse covariance matrix approach as an alternative to full Bayesian inverse Wishart simulation.

Bayesian methods have become increasingly popular as a solution to the over-parameterization in VAR (vector autoregression) and SUR (seemingly unrelated regression) systems. An SUR model consists of a set of regressions that may seem unrelated but have correlated error terms. A VAR(p) system is an SUR model where each equation uses the same regressors. To learn about interaction effects among multiple shocks, we employ a Lasso-tilted inverse Wishart prior that forces sparsity in the variance-covariance matrix estimation. Our methodology jointly regularizes expected values and variance-covariance matrices in VAR and SUR systems in a computationally efficient way. Particular attention is paid to the sensitivity over a wide range of hyper-parameters.

Our method reinterprets Bayesian studies of equity-premium predictability of \cite{Kandel1996} and \cite{Barberis2000}. We find a plausible prior specification for predictability in S\&P 500. In a second study, we provide a sensitivity analysis for the bond risk premia prediction using macro factors, see \cite{ludvigson2009macro, ludvigson2010factor}. Our regularized SUR prediction, with a multi-response perspective, uses supervised learning to explore the common macro factors in excess returns for bonds of multiple maturities. Our fast prediction sensitivity check is an alternative to a typical MCMC sampling approach.

Rather than marginalizing out prior hyper-parameters in a fully Bayesian analysis, we calibrate hyper-parameters using an economically meaningful predictive cross-validation measure given a univariate target. We develop a proximal algorithm using majorize-minimize operations (see \cite{polson2015proximal} and \cite{Bien2011}) to provide a simple alternative to MCMC simulation for high-dimensional VAR and SUR systems. In addition to computation speed, both empirical studies show that our proposed method is very competitive in terms of predictor selection, as well as prediction accuracy.

The rest of the paper is outlined as follows. Section \ref{sec11} provides the intuition behind the connection between prior sensitivity analysis and model regularization, and Section \ref{sec12} adds a literature review. Section \ref{sec2} presents the regularized system for Bayesian VAR and SUR, as well as a discussion of the Bayesian MAP estimator. Section \ref{sec3} develops a proximal algorithm for both auto-regressive coefficients and variance-covariance matrices within a VAR setting. Section \ref{sec41} illustrates our method by revisiting the equity-premium analysis for return predictability using many economic fundamental predictors from \cite{Welch2008}. Section \ref{sec42} revisits a well-known study about bond prediction using a high-dimension of macro factors from \cite{ludvigson2009macro,ludvigson2010factor} and provides insights beyond those provided by the dynamic factor analysis. Finally, Section \ref{sec5} concludes with directions for future research.

\subsection{Prior Sensitivity and Regularization \label{sec11}}
To fix notation, define $B$ as an auto-regressive coefficient in a VAR(1) model, and $\Sigma$ as a variance-covariance matrix for multiple shocks. Statistical regularization requires a researcher to specify a measure of goodness of fit, denoted by $l(B, \Sigma)$, as well as a penalty function that achieves a parsimonious model, denoted by $ \phi(B, \Sigma)$. Probabilistically,  $l(B, \Sigma)$ and $\phi(B, \Sigma)$ correspond to the negative logarithms of the likelihood and a prior distribution. Regularization leads to an optimization problem of the form.
\[ \underset{B, \Sigma \in \Re^d}{\min} l(B, \Sigma) + \phi(B, \Sigma). \]
For example, a regularized regression minimizes a least-squares objective (Gaussian likelihood) plus a penalty such as an $L^2$-norm (Ridge) Gaussian probability model or $L^1$-norm (Lasso) double exponential probability model. 

A probabilistic approach, on the other hand, leads to a Bayesian hierarchical model
\[ p(y \mid B, \Sigma) \propto \exp\{-l(B, \Sigma)\}, \quad p(B, \Sigma) \propto \exp\{ -\phi(B, \Sigma) \}.\]
The solution to the minimization problem corresponds to maximizing the posterior density, 
\[ p( B, \Sigma \mid y) \propto p(y \mid B, \Sigma)\times p(B, \Sigma) = \exp\{- l(B, \Sigma) - \phi(B, \Sigma)\} \]
\[ (\hat{B}, \hat{\Sigma}) = \argmax_{B, \Sigma} \; p( B, \Sigma \mid y), \] where $ p(B, \Sigma \mid y)$ denotes the posterior distribution. Here, $(\hat{B}, \hat{\Sigma})$ is simply the posterior mode.

Under a decomposable penalty, $\phi(B, \Sigma) = \lambda\phi_1(B) + \gamma\phi_2(\Sigma) $, the $\lambda$ and $\gamma$ hyper-parameters of a prior distribution are tuning parameters in a regularization problem. Whereas a Bayesian study requires a prior distribution and is sensitive to its hyper-parameters $(\lambda, \gamma)$, a regularization problem uses $(\lambda, \gamma)$ to control the bias-variance tradeoff of the model complexity. Consequently, the regularized estimates $(\hat{B}, \hat{\Sigma})_{(\lambda,\gamma)}$ provide a regularization path, which can be interpreted as prior sensitivity analysis for the MAP estimator. Therefore, the pair of $\lambda$ and $\gamma$ is the key to connect the interpretation of prior sensitivity analysis and model regularization. 

\subsection{Connection with Empirical Macro-Finance \label{sec12}}

\cite{Campbell1989} and \cite{Fama1988} provide early evidence of time-series predictability of stock returns and show that market returns can be predicted using lagged dividend yields. However, \cite{Welch2008} examine 14 predictor variables and find little forecasting power in univariate forecasting regressions. \cite{Cochrane2008} uses a VAR system of returns and dividend growth to explore their joint stochastic dynamics and defend the return predictability.  \cite{Campbell2008} find an economically significant out-of-sample forecasting power after imposing economically reasonable parameter restrictions. Therefore, we build on the empirical finance literature by adopting a regularized Bayesian predictive framework to exploit the joint dynamics for the univariate predictor and provide a forecast sensitivity analysis. 

Early studies in time-varying bond risk premia include \cite{fama1987information} and \cite{campbell1991yield} regarding forecasting yield changes with yield spreads. The excess returns on U.S. government bonds can be predictable by the cross-section of yield spreads or forward rates. \cite{cochrane2005bond} later propose a return-forecasting factor predicts excess returns on one- to five-year maturity bonds with R$^2$ up to 44\%. From a large number of macro series, \cite{ludvigson2009macro} use of principal components analysis to estimate common factors regarding bond prediction. \cite{ludvigson2010factor} use prior information to organize the macro series into 8 subgroups and estimate a dynamic factor model for each subgroup using a Bayesian estimation. 

\cite{giannone2015prior} analyze the hyper-parameter uncertainty for the density forecasts in a VAR model. With predictor selection in a large VAR to reduce estimation uncertainty, they use informative priors to shrink the richly parameterized unrestricted model and reduce prediction errors. Whereas \cite{carriero2012forecasting} study a large BVAR and optimally select the amount of shrinkage by maximizing the marginal likelihood, our approach relies on a predictive data-driven selection. \cite{banbura2010large} also apply cross-validation to estimate one single hyper-parameter in a large VAR, but their approach performs neither variable selection nor variance-covariance regularization.

\cite{stock2002forecasting, Stock2002} find significant improvements for macro and financial predictions using common factors estimated from large data sets. \cite{jurado2015measuring} attempt to reduce the dimension of macro series to quantify the time-varying macroeconomic uncertainty. For the recent development in high-dimensional time-series models, \cite{chan2016large} propose a Bayesian approach for inference in VARMA, and \cite{Nicholson2015} introduce regularization to reduce the parameter space of VAR and VARX (VAR with exogenous variables) models. \cite{Zantedeschi2011} also develop a Bayesian procedure for macro-finance forecasting.

Given the extensive list of potential predictors for both market risk premium and bond risk premium, we consider the role of shrinkage priors to examine the model uncertainty about the existence and strength of predictors. \cite{Park2008} suggest using the posterior mode interpretation of Lasso regularization. \cite{xu2015bayesian} propose the posterior median estimator for the Bayesian group lasso and uses spike and slab priors for group variable selection. In a hierarchical Bayesian framework, \cite{yuan2005efficient} show the empirical Bayes estimator that is closely related to the Lasso estimator for variable selection.

\section{Regularizing Bayesian Predictive Regressions \label{sec2}}

To illustrate Bayesian SUR system, we discuss high-dimensional VAR regularization which is popular in the empirical macro-finance literature. We demonstrate our SUR regularization in the bond prediction in Section \ref{sec42}. 

\subsection{Bayesian Seemingly Unrelated Regressions} 

 \cite{Zellner1962, zellner1963estimators} initially proposed SUR to estimate a system of stacked regression equations, which include cross-equation parameter restrictions and correlated error terms. The model is also referred to as a ``generalized multivariate regression model" and therefore can be solved in a generalized least squares approach. Specifically, we have a matrix formulation for a general SUR model of the form
\begin{equation}
\label{eqn:SUR}
    Z = XB + E, \quad \mbox{where } E \sim \mathcal{N}(0, \Sigma \otimes I). 
\end{equation} 
Here the target Z is an $n\times m$ matrix. $\mathcal{N}(0, \Sigma \otimes I)$ denotes the multivariate normal distribution and $\otimes$ is the Kronecker product. $\Sigma$ is an $m\times m$ matrix. Throughout, we use the following stacked variables $Z^\intercal = \bmat{Z_1^\intercal, \cdots, Z_m^\intercal}$, $X = \mbox{diag}\{X_1, \cdots, X_m\}$, $B^\intercal = \bmat{\beta_1^\intercal, \cdots, \beta_m^\intercal}$, and $E^\intercal = \bmat{\epsilon_1^\intercal, \cdots, \epsilon_m^\intercal}$.

The regressor $X_i$ for each individual regression is a $T \times p$ matrix and $\beta_i$ is a $p \times 1$ vector. The likelihood function is
\begin{eqnarray*}
	L(Z, X \mid B, \Sigma) &=& (2\pi)^{-nm/2} \det(\Sigma)^{-n/2} \exp\left(-\frac{1}{2}(Z - XB)^\intercal (\Sigma^{-1} \otimes I) (Z - XB) \right) \\
	&=& (2\pi)^{-nm/2} \det(\Sigma)^{-n/2} \exp\left(-\frac{1}{2}\mbox{tr}\{S_B\Sigma^{-1}\}\right),
\end{eqnarray*}
where the (i, j)th element in $S_B$ equals to $(Z_i - X_i\beta_i)^\intercal (Z_j- X_j\beta_j)$. Given a non-informative Jeffreys' invariant prior, the joint posterior density function is 
\[ p(B, \Sigma \mid Z, X) \propto \det(\Sigma)^{-(n+m+1)/2}\exp\left(-\frac{1}{2}\mbox{tr}\{S_B\Sigma^{-1}\}\right) \]

\cite{Zellner1971} introduces a Bayesian approach to calculating posterior densities for parameters to estimate the SUR model. In the recent advancement, \cite{Zellner2010} propose a direct Monte Carlo (DMC) approach to calculate Bayesian estimation and prediction results using diffuse or informative priors. \cite{rothman2010sparse} propose a penalized likelihood method with simultaneous estimation of the regression coefficients as well as the covariance structure. \cite{chen2012sparse} apply a group-lasso type penalty in predicting multiple response variables from the same set of predictor variables. 

\subsection{Vector Auto-Regressions}

The class of VAR models are a popular forecasting tool for empirical financial and macroeconomic time-series analysis that can capture complex dynamic interrelationships among variables. \cite{Kandel1996} adapt a VAR formulation to investigate the predictability of the equity premium and build portfolios using the predictive return distribution. Our predictive cross-validation in Section \ref{sec33} follows their framework of univariate variable forecasting in a multivariate time-series model.

In general, a VAR(p) system jointly explores the stochastic dynamics of both stock market returns, denoted by $y_t$, and economic predictive variables, $x_t$, for $1\leq t \leq T$, where
\begin{equation*}
\bmat{y_t \\ x_t}  = \alpha + \beta_1\bmat{y_{t-1} \\ x_{t-1}} + \cdots + \beta_p\bmat{y_{t-p} \\ x_{t-p}} + \epsilon_t, \quad \mbox{ where } \epsilon_t \sim \mathcal{N}(0, \Sigma).
\end{equation*}
Here $y_t$ is a continuously compounded excess market return, and $x_t$ is a vector of $K$ economic predictive variables, which typically include dividend yield, earning-price ratio, book-to-market ratio, and so on. For multi-step prediction, we can estimate the VAR parameters $(\alpha, \beta, \Sigma)$ and iterate the model forward with the parameters fixed at their estimated values.	

To illustrate the model, consider the simplest case of a VAR(1) system. Given that most predictors proposed in the literature are lagged one period, we rewrite a VAR(1) model with a demeaned vector given by $Z_t = (y_t, x_t^\intercal)^\intercal - E_T[(y_t, x_t^\intercal)^\intercal]$ and the autoregressive structure:
\begin{equation} \label{VAR1}
	Z_t = B Z_{t-1} + \epsilon_t 
\end{equation}

In our empirical study of Section \ref{sec41}, the goal is to study the joint predictability of proposed predictors to stock returns and see the trade-off between model specification and forecasting power. To compute the posterior distribution, it is convenient to reformulate the VAR(1) model into the matrix formulation of SUR in (\ref{eqn:SUR}), and we write
$z = \mbox{Vec}(Z) = (L^\intercal \otimes I_{p})B+ \mbox{Vec}(E)$, where Vec($\cdot$) is the column stack operator, and we stack variables and parameters $Z = (Z_1, \cdots, Z_{T})$, $L = (Z_0, Z_1, \cdots, Z_{T-1})$, $E = (\epsilon_1, \epsilon_2, \cdots, \epsilon_{T})$ and $B = \mbox{Vec}(\beta)$. 

\subsection{Bayesian MAP Estimator}

We now turn to the problem of computing a regularized Bayesian MAP (Maximum-a-Posteriori) estimation. With a joint penalty, denoted by $\phi(B, \Sigma)$, for the parameters $\beta$ and the variance-covariance matrix $\Sigma$, a MAP estimator corresponds to the mode of the posterior distribution. The usual Bayesian estimator is the posterior expectation that minimizes the quadratic loss in Bayesian decision studies, whereas the MAP estimator minimizes the 0-1 loss. When the prior density is flat, the posterior mode turns out to be the maximum likelihood estimator.

This problem is equivalent to solving a penalized likelihood with regularization where $\phi(B, \Sigma)$ is considered a penalty function to control the favorable bias-variance trade-off. \cite{polson2015mixtures} describe such a duality between specifying a regularization penalty and a prior distribution. The prior hyper-parameter $\lambda$ can be viewed as the amount of regularization. We demonstrate that using some shrinkage priors performs variable selection, whereas $\lambda$ is used to control the model size and assess the bias-variance trade-off.

\cite{Kandel1996} study the equity-premium puzzle under a Gaussian prior for B, which is equivalent to a shrinkage $L^2$ penalty, where $B \sim \mathcal{N}(0, \tau^2 I) \mbox{ and } p(B) \propto \exp\Big(-\frac{B^\intercal B}{2\tau^2}\Big)$. With a Jeffrey's prior, we set $p(\Sigma) = 1$ and then can obtain the $L^2$ regularized MAP posterior mode estimate to compare the Bayesian posterior mean estimate of Kandel and Stambaugh by finding 
\[ B^\star = \argmin_{B} l(B \mid \Sigma) + \lambda\lVert B\rVert_2^2. \]

These penalized model comparison criteria is shown to correspond to a hierarchical Bayes model selection procedure under a particular class of priors by \cite{george2000calibration}. The process of selecting predictor variables can then be simplified using Laplace shrinkage priors, where $B_i \iid \mbox{Laplace}(0, \lambda) \mbox{ with } \phi(B) \propto \exp\{-\lambda\sum_{i=1}^P\lvert B_i \rvert \}$. An $L^1$ penalty performs variable selection through the sparsity of B and the optimization problem $ B^\star = \argmin_{B}\Big\{ l(B \mid \Sigma) + \lambda\lVert B\rVert_1\Big\}$. Such a shrinkage prior is a solution to the unstable inference and inaccurate out-of-sample forecasts in large (Bayesian) VARs with dense parameterization. 

\subsection{Prior Distribution of B and $\Sigma$}

For regularizing large variance-covariance matrices, \cite{Das2010} also present a generalized multivariate gamma distribution and discuss the MAP covariance estimation. \cite{cai2011adaptive} consider estimation of sparse covariance matrices and propose an adaptive thresholding procedure that outperforms the universal thresholding estimators. \cite{fan2013large} introduce the principal orthogonal complement thresholding method to deal with a sparse error covariance matrix in an approximate factor model. 

When estimating a covariance matrix, $\Sigma$, two common prior distributions are Jeffreys' prior and the inverse Wishart. We now wish to model sparsity in the elements of the correlation matrix, denoted by $\rho$, as a sparse structure. The standard prior distribution for a variance-covariance matrix is the family of inverse Wishart probability densities, $p_{IW}(\Sigma)$, given by
\[ p_{IW}( \Sigma \mid v, S_0 ) \propto \det(\Sigma)^{\nu+p/2}\exp \left\{ - \frac{1}{2} \mbox{tr}( S_0 \Sigma^{-1} ) \right\}. \]

\noindent Here $ (S_0,\nu)$ are prior hyper-parameters. There are a number of approaches for specifying these parameters. For example, Jeffrey's prior corresponds to the special case $p(\Sigma) \equiv det(\Sigma)^{p/2}$. Zeros in a covariance matrix correspond to marginal independencies between variables and are desirable for impulse-response functions in VAR analysis. We will achieve this estimation by imposing an $L^1$ regularization penalty.

One of the main advantages of using an $L^1$-penalty is that the MAP estimator, $\Sigma^\star$, has zeros forced with its solution. Modeling $L^1$-sparsity in $\Sigma$ corresponds to adding an appropriate scale matrix and hyper-parameters for degrees of freedom to get a so-called Lasso-tilted inverse Wishart density, denoted by $p_{LIW}(\Sigma)$, where
\begin{equation}
p_{LIW}(\Sigma \mid B, Z) \propto \det(\Sigma)^{v+p/2}\exp\{-\mbox{tr}(S_0\Sigma^{-1})\}\exp\{-\gamma \Vert \Sigma \Vert_1\}.
\end{equation}
Jointly regularizing $(B, \Sigma)$ in a VAR system leads to a posterior density with $S_B = \frac{1}{T} \sum_{t=1}^T \epsilon_t \epsilon_t^\intercal$,
\[ p_{LIW}(\Sigma \mid B, Z, Y) \propto \det(\Sigma)^{v+(T+p)/2}\exp\{-\mbox{tr}((S_B+S_0)\Sigma^{-1})\}\exp\{-\gamma \Vert \Sigma \Vert_1\}. \]

\cite{Bien2011} provide convex optimization with an $L^1$-penalty can be applied to a variance-covariance matrix by penalizing the entries of the covariance matrix. Specifically, they show how to minimize a penalized negative log likelihood
\begin{equation}
\label{BienOptim} 
\Sigma^\star = \argmin_{\Sigma}\Big\{ \log(\det(\Sigma)) + \mbox{tr}(S \Sigma^{-1}) + \gamma \Vert \Sigma \Vert_1 \Big\}, 
\end{equation} 
where the $L^1$ norm definition for a matrix is $\lVert \Sigma \rVert_1 = \lVert \mbox{vec}(\Sigma) \rVert_1 = \sum_{i,j}\lvert \sigma_{i,j}\rvert$. \cite{Bien2011} use a ``Majorize-Minimization" (MM) approach to solve the minimization problem and provide a \texttt{R} package [\texttt{spcov}]. We show how to adapt their optimization solution into our joint regularization problem in Section \ref{sec32}. 

\section{Regularizing VARs \label{sec3}}

The sparse Lasso-VAR and shrinkage Ridge-VAR are identical in model reformulation and penalized estimation via coordinate descent algorithms. Both of them can be reformulated to an elastic net regularization, which linearly combines the $L^1$ and $L^2$ penalties of the Lasso and Ridge methods. Regularization produces a convenient tool to calculate posterior modes and graphically display prior sensitivity analysis via a regularization path. The model setup below is based on Ridge-VAR, and empirical results of Ridge-VAR are provided in Section \ref{sec41} and \ref{sec42}.

A high-dimensional Bayesian VAR is computationally convenient and interpretable by shrinkage regularization. However, \cite{Song2011} show the negative consequences of directly applying Lasso-type penalties for time series without considering the temporal dependence. Then \cite{Davis2015} use a maximum likelihood approach and propose a two-stage approach for fitting sparse VAR. Their procedure is based on the property that, for a given variance-covariance matrix, the penalized likelihood can be recast into a penalized regression. 

The idea is the same as stacking equations to solve these simultaneous systems, such as VAR and SUR. The likelihood estimation of Lasso-VAR models is not straightforward, because the likelihood function involves the unknown parameter $\Sigma$ and estimation of the variance-covariance matrix $\Sigma$ is usually difficult because of the curse of dimensionality. We have adapted their likelihood reformulation into our regularized VAR framework.

\subsection{Regularization of B and $\Sigma$}
Given a sequence of $T$ multivariate normal $p$-dimensional random vectors $\{\epsilon_t\}$, and then the negative log likelihood using $B$ and $\Sigma$ is
\[ l(B, \Sigma) = \frac{T\times p}{2}\log 2\pi + \frac{T}{2}\log\{\det(\Sigma)\} + \frac{T}{2}\mbox{tr}(\Sigma^{-1}S_B). \]
\noindent where $S_B = \frac{1}{T}\sum_1^{T}\epsilon_t\epsilon_{t}^\intercal$ and $\epsilon_t = Z_t - B Z_{t-1}$.

Given our model specification and stack variable notations in (\ref{VAR1}), the conditional negative log-likelihood $l(B \mid \Sigma)$ of the VAR(1)  model is given by
\[ l(B \mid \Sigma) \propto  T\log\{det(\Sigma)\} + [z - (L^\intercal \otimes I_{p})B]^\intercal(I_T \otimes \Sigma^{-1})[z - (L^\intercal\otimes I_{p})B]. \]

\noindent The negative log-likelihood of $\Sigma$ conditional on B is given by
\[ l(\Sigma \mid B) \propto \log\{\det\Sigma\} + \mbox{tr}(\Sigma^{-1}S_B). \]

To regularize B, we add the penalty function on $ l(B \mid \Sigma)$ and calculate the estimate $B^\star = \argmin_{B} \Big\{ l(B \mid \Sigma) + \lambda\lVert B \rVert_1 \Big\}$. Maximizing such a penalized Gaussian likelihood is equivalent to minimizing a penalized least-squares errors. Therefore, the optimization problem can be efficiently computed using the \texttt{R}  package [\texttt{glmnet}]. Similarly, for the Ridge regularization for a Gaussian prior specification, we can solve $B^\star = \argmin_{B} \Big\{l(B \mid \Sigma) + \lambda\lVert B \rVert_2\Big\}$. 

Similarly, we add a scaled $L^1$-penalty function on $ l(\Sigma \mid B)$ for the joint regularization in ($B, \Sigma$ ). But we need to solve the following optimization problem, where $\Sigma \succ 0$ means positive definiteness for covariance matrix,
\begin{equation} 
\argmax_{B, \Sigma} p(B, \Sigma \mid y) = \argmin_{\Sigma \succ 0} \{l(\Sigma \mid B) + \psi(B) + \gamma\lVert P \odot \Sigma \rVert_1\},
\end{equation}

\noindent where $P \odot \Sigma_\epsilon$ is defined as element-to element multiplication and P is a $p\times p$ matrix of all 1's with 0's on the diagonal to ensure the positive-definiteness. When using an infinite penalty $\gamma$,  the solution is a diagonal variance-covariance matrix with all zero covariance pairs. The optimization algorithm is allowed to regularized variance-covariance matrix, including the level of variance in the diagonal, but we only regularize the correlation matrix in the empirical analysis.

\subsection{Proximal Algorithm \label{sec32}}

The goal is to calculate the mode of the posterior distribution, $\argmax_{B, \Sigma} p(B, \Sigma \mid y)$. Although the objective function in (\ref{BienOptim}) is not convex, it is the sum of a convex function and a concave function. We can see $tr(S\Sigma^{-1}) + \gamma\lVert \Sigma \rVert_1$ is convex in $\Sigma$, but $\log(\det(\Sigma))$ is concave. Therefore, we adapt the MM algorithm suggested in \cite{Bien2011} to solve this optimization for $(B^\star, \Sigma^\star)$. In short, the MM method is a prescription for constructing interactive optimization algorithms that exploit the convexity of a function to find the minima. 

We now combine these two convex programming problems to construct a proximal algorithm that jointly regularizes B and $\Sigma$. We have a composite objective of the negative log-likelihood function, regularization of B and regularization of $\Sigma$.  Our joint objective depends on two global regularization parameters $(\lambda, \gamma) >0$ and is specified by
\begin{equation} 
(B^\star, \Sigma^\star) = \argmin_{\Sigma \succ 0, B} \Big\{ Q(B,\Sigma) = l(B,\Sigma) + \lambda\lVert B \rVert_1 + \gamma\lVert P \odot \Sigma \rVert_1 \Big\}
\end{equation}
\noindent by iterating the optimal parameters of B and $\Sigma$ from the individual optimizations.
\[ B^{(k+1)} = \argmin_{B} Q(B \mid \Sigma^{(k)}) = l(B,\Sigma) + \lambda\lVert B \rVert_1 \]
\[ \Sigma^{(k+1)} = \argmin_{\Sigma \succ 0} Q(\Sigma \mid B^{(k+1)}) = l(B,\Sigma) + \gamma\lVert P \odot \Sigma \rVert_1 \]

Hence, we have constructed a sequence $\{B^{(k+1)}, \Sigma^{(k+1)} \}$ that converges to the Bayesian MAP estimator $(B^\star, \Sigma^\star)$. As we vary the regularization parameters $(\lambda, \gamma)$, we trace out a full regularization solution path, thereby providing our sensitivity diagnostics. \cite{polson2015proximal} provides a discussion of convergence properties of proximal-point algorithms and shows how gains in efficiency can be achieved with Nesterov's acceleration.

One advantage of our joint regularization is that it is computationally efficient because we divide the composite objective and conquer each optimization sequentially. In the empirical analysis, we need no more than ten iterations for a sufficient convergence. The empirical results with a full prior sensitivity analysis are especially useful in empirical finance and macroeconomics forecasting. For example, we can learn about variable selection via increasing sparsity in B, and how target-predictor impulse-responses vary through to increasing the sparse correlation $\rho$ of $\Sigma$. 

\subsection{Predictive Cross-Validation \label{sec33}}
Given the duality between the pair of tuning parameters $(\lambda, \gamma)$ with the hyper-parameters of a prior distribution of $(\beta, \Sigma)$, we can use predictive cross-validation to help select optimal prior hyper-parameters rather than directly marginalizing them out. The goal of out-of-sample prediction is to show regularization paths of prediction error as a function of tuning parameters. From the perspective of predictive ability, one can search the optimal pair of $(\lambda, \gamma)$ to achieve the best out-of-sample performance in a hold-out sample of data. 

If our goal is the one-step-ahead prediction of the equity premium, we can apply a predictive cross-validation to calculate the model-prediction error in history. Cross-validation is an intuitively data-driven resampling method to assess the model out-of-sample performance and is extremely useful for model selection. We can then estimate a sequence of VAR models by using a rolling window of data and obtain the one-step-ahead prediction error. \cite{Song2011} also suggest choosing tuning parameters via a data-driven rolling scheme method to optimize the forecasting performance. For a robustness check, we have also implemented AIC and BIC for tuning-parameter selections. For the market-timing strategy in Section \ref{sec413}, the AIC selection demonstrates strong predictability in the multiple-predictor model.

Our equity-premium examples in Section \ref{sec41} show 63 years of quarterly data, and hence we pick the window size as 20 years or 80 quarters. Eighty observations is still an adequate number for estimating a regular 10-dimensional VAR(1) model. We have 172 overlapping rolling schemes in the sample, which is the number of VAR model estimation. The model-comparison criteria are the sample mean of the prediction errors in S\&P 500 excess returns. Other variables in the VAR system are viewed as ``instruments" in the forecast. Therefore, we can find a way to determine the level of regularization with the best out-of-sample prediction performance. 

For multi-step-ahead forecasts, we can change the objective function to minimize the multi-step-ahead prediction errors. In this scenario, our proposed data-driven selection method is more ``informative" than any prior specification that is only related to one-step-ahead forecasts. Our VAR formulation could be applied to a multi-dimension forecast by using the rest of variables as ``predictive instruments" in the time series system. In the excess bond return prediction example of Section \ref{sec42}, we only show the sensitivity analysis, but we can use the same predictive cross-validation to determine the best tuning parameters. The advantage for the supervised learning in a multiple-response model is to specify the common tuning parameter. 

In short, predictive cross-validation provides a powerful connection between a regularizing predictive regression and a fully Bayesian approach. Traditionally, we specify the prior hyper-parameters and calculate out-of-sample predictions, but this approach is optimal only under some specific priors. Our regularization approach demonstrates exactly how different prior hyper-parameters affect predictive power through regularization plots. Specifying one data-driven prior in the Bayesian predictive system is possible. We suggest one can view a regularization approach as a quick precursor to a more detailed full Bayesian analysis.

\section{Regularizing Macro-Finance Predictions}

\subsection{Equity-Premium Prediction \label{sec41}}
To illustrate our methodology, we revisit the equity-premium predictors surveyed in \cite{Welch2008}, who argue all these predictors lack out-of-sample forecastability. We examine the robustness of their predictability in both single-predictor and multiple-predictor models. For Bayesian VAR studies, \cite{Kandel1996} provide a framework of Bayesian predictive regressions using multiple predictors and a zero-mean prior that implies no time-series predictability and market efficiency. \cite{Barberis2000} studies a single-predictor model using the same conventional non-informative prior. Moreover, we examine the sensitivity of the evidence on predictability through a market-timing strategy as one out-of-sample study: whether investors can exploit the predictability and earn profits more than a buy-and-hold strategy in the market index. 

\subsubsection{Single-Predictor Model}
In our analysis, the excess market return is the quarterly return on the S\&P 500 index minus the short-term risk-free rate. Table \ref{Tab0} has a description of the 14 economic fundamental variables proposed by academics. Our analysis uses quarterly data for excess returns and the dividend-price ratio of the S\&P 500 index from 1952 to 2015. We use a Gaussian prior for the AR coefficient and the Lasso-tilted inverse Wishart prior for the variance-covariance matrix. The optimal amount of regularization on B and $\Sigma$ is calibrated to the least predictive MSE for one-step-ahead excess returns.  

\cite{Barberis2000} uses a ``non-informative" prior that is equivalent to the case of $\lambda=1$ for B and $\gamma=0$ for $\Sigma$ in our regularization. Table \ref{Tab1} shows the estimation comparison between Barberis's Bayesian posterior mean approach and our corresponding MAP mode approach. We can compare Barberis's coefficient B and our $\beta_{[:, 2]}$, which are coefficients of (D/P)$_{t-1}$ on return$_t$ and (D/P)$_{t}$. First, the dividend-price ratio is a random walk in a full Bayes estimation and must underperform the regularized estimator. Second, the large difference in $\Sigma_{[2, 2]}$ is due to this random-walk specification, and our only focus is the excess-return variable. Finally, The amount of optimal regularization suggests the predictability of the dividend-price ratio should be nonzero but slightly lower.

\begin{figure} 
\caption{Regularizing Single-Predictor Model.\label{Fig1}}
\vspace{0.2in}
\includegraphics[width=\textwidth]{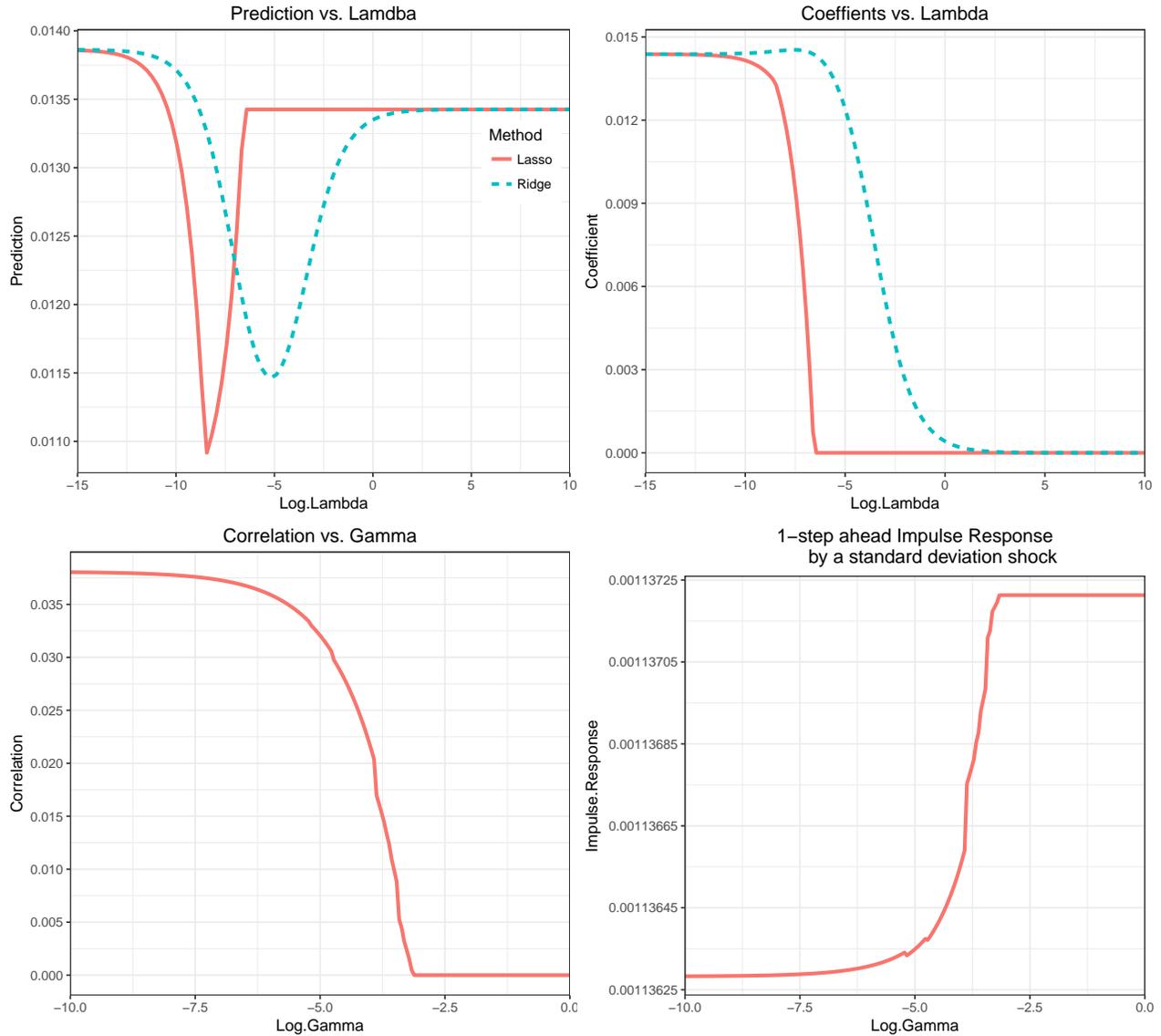}
\vspace{0.2in}

\small The top two plots show regularization path of regularizing the AR coefficients when no regularization is in the variance-covariance matrix. The top left panel is about prediction errors. The top right panel shows coefficient estimates for Dividend-Price Ratio. The bottom two plots show regularization path of regularizing variance-covariance matrix when no regularization is on the AR coefficients. The bottom left panel is about correlation for the variance-covariance matrix. The bottom right panel plots 1-step ahead orthogonalized impulse response.
\end{figure}

Figure \ref{Fig1} plots summaries for the single-predictor model from 1952 Quarter 1 to 2015 Quarter 4. To isolate the regularization effect, we regularize the AR coefficients for the top two plots and the variance-covariance matrix for the bottom two plots, respectively, in Figure \ref{Fig1}. The top-left plot shows how the predicted values vary over a wide range of tuning parameters. We see sequentially incorporating the belief of predictability dramatically changes the predicted value of S\&P 500 excess returns for 2015 Quarter 4. The shrinkage priors allow us to observe the signal strength sequentially through regularization. The minimum values of prediction regularization are in the middle and are different in Ridge and Lasso shrinkages. The left end corresponds to a least regularized prediction (VAR), and the right end corresponds to a most regularized prediction (Average). 

Figure \ref{Fig1} bottom-left panel shows the regularization path of correlations, and the bottom-right panel shows the regularization path of one-step-ahead orthogonalized impulse response from the dividend-price ratio to S\&P 500 excess returns.  When we increase the shrinkage amount $\gamma$, we can see the shock correlation shrinks to zero, but the orthogonal impulse response hardly changes (see the y-axis values). The advantage of using an orthogonal impulse response is the consideration of the shock relationship. But we find regularizing $\Sigma$ has negligible effects on the cross-impulse-response function from the dividend-price ratio to excess returns.

\subsubsection{Multiple-Predictor Mode}
Many previous discoveries of economic predictors for the equity premium are in the univariate forecasting model. \cite{Welch2008} perform their influential study with individual predictors instead of a combination of all predictors, and we revisit their proposed predictors at the lens of our high-dimensional regularized VAR. However, \cite{Rapach2010} find combinations of different model forecasts outperform the historical average on a consistent basis over time. Variable selection is an additional insight that requires spike-and-slab priors (see \cite{ishwaran2005spike}) within the Bayesian MCMC. Using the shrinkage regularization, we can learn about the prior influence at the regularization-parameter cutoffs where variable selection occurs. 

\begin{figure}
\caption{Regularizing Multiple-Predictor Model.\label{Fig2}}
\vspace{0.2in}
\includegraphics[width=\textwidth]{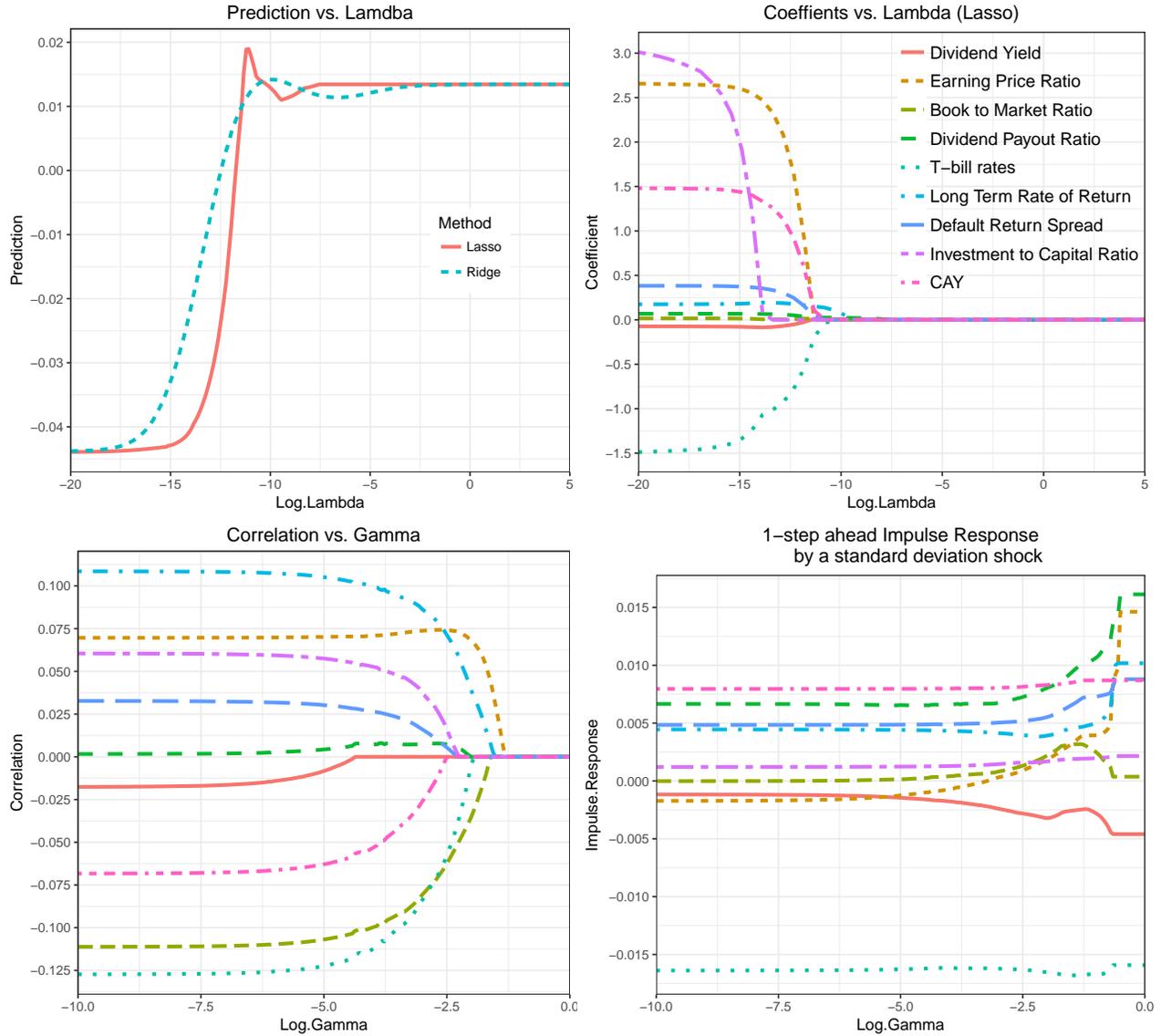}
\vspace{0.2in}

\small The top two plots show regularization path of regularizing the AR coefficients when no regularization is in the variance-covariance matrix. The top left panel is about prediction errors. The top right panel shows coefficient estimates for multiple predictors using the Lasso shrinkage prior. The legend of all predictors is located in the bottom left panel. The bottom two plots show regularization path of regularizing the variance-covariance matrix when no regularization is on the AR coefficients. The bottom left panel is about correlations for the variance-covariance matrix. The bottom right panel plots 1-step ahead orthogonalized impulse response.
\end{figure}

We explore the equity-premium predictability with many economic variables in a joint dynamics from 1952 Quarter 1 to 2015 Quarter 4. Figure \ref{Fig2} plots the estimation summaries of the multiple-predictor model and illustrates the power of regularization for high-dimensional data. Similarly, we regularize the AR coefficients for the top two plots and the variance-covariance matrix for the bottom two plots, respectively. The top-left plot shows how the predicted values vary over a wide range of tuning parameters. We see sequentially incorporating the belief of predictability lowers the predicted value of S\&P 500 returns for 2015 Quarter 4 from positive to negative. The top-right plot shows the changes in predictor existence and the strength for the predictor excess stock returns. The property of sparsity allows us to observe the signal strength sequentially through regularization. We find the top three predictors are the book-to-market ratio, CAY (consumption, wealth, income ratio),  and T-bill rates. 

The bottom two plots show a regularization path for regularizing the variance-covariance matrix without regularization on the AR coefficients. The bottom-left panel shows correlations between predictors and returns for the covariance matrix. The top three predictors that have the most correlated shock with returns are the book-to-market ratio, earning-price ratio, and long-term rate of return. The bottom-right panel plots the one-step-ahead orthogonalized impulse response from predictors to returns. The top three predictors that have largest impulse response to returns are dividend yield, dividend payout ratio, and earnings-price ratio. Also, given that all impulse-response plots are flat, the uncorrelated-shock assumption does not affect the study of the impulse response.

\subsubsection{Market-Timing Strategy \label{sec413}}
\cite{Samuelson1969} and \cite{Merton1969} find an investor with power utility wants to hold a fixed portfolio of stocks and bonds when facing i.i.d. Market returns. If the excess market return is predictable, one can perform a market-timing strategy to change his fixed portfolio split. Therefore, implementing a market-timing strategy can test the out-of-sample performance of our regularization strategies over the buy-and-hold strategy. We also provide three strategies to check the robustness: the most regularized forecast (historical moving average), the optimally regularized one-step-ahead forecast, and the least regularized VAR(1) forecast.

Admittedly, all return forecasts should be considered pseudo-out-of-sample in our analysis, because we recursively re-estimate the model and then perform prediction. Previous studies about market-timing strategies do not consider the impact of taxes but might consider a fixed transaction cost. We consider neither of the tax impact and transaction cost in our analysis because we only update the portfolio quarterly and limit the portfolio change to no more than 50\%. For example, I have a portfolio with value \$1. The maximum change we can make in this quarter is either sell \$0.5 S\&P 500 and buy \$1 risk-free rate or sell \$0.5 risk-free rate and buy \$1 S\&P 500. Also, we restrict short selling in this simple exercise.

In the market-timing exercise, we look for the one-step-ahead market-return forecast and update the mean-variance optimal portfolio between the stock and risk-free rate every quarter. To avoid look-ahead bias, we fix the optimal regularization tuning parameter that is estimated using the training data from 1970 Quarter 1 to 1989 Quarter 4. Therefore, setting the optimal regularization amounts for ($\gamma$, $\lambda$) is one way to perform out-of-sample testing. For every quarter from 1990 to 2015 in the testing data, we use a moving training period of the previous 20 years (80 quarters) to re-estimate the regularized model and obtain the one-quarter ahead forecast for equity premium. Therefore, parameters are then re-estimated for each forecasting model, and the market-timing strategy is performed recursively. 

With a maximum 50\% monthly portfolio change, we update the stock-bond split to the mean-variance optimal level quarterly based on the model prediction and sample variances. We do not model the portfolio variance but calculate the sample variance using the same training period (80 quarters). The maximum 50\% change might be an unrealistic assumption, but our purpose is to adapt a smooth loss function without considering transaction cost to compare the performances of three approaches. We also compare all three strategies with the buy-and-hold strategy, the fixed-split portfolio, with the same starting stock and bond shares.

\begin{figure}
\caption{Market Timing Strategy (Predictive Cross-Validation) \label{Fig3a}}
\vspace{0.2in}
\includegraphics[width=\textwidth]{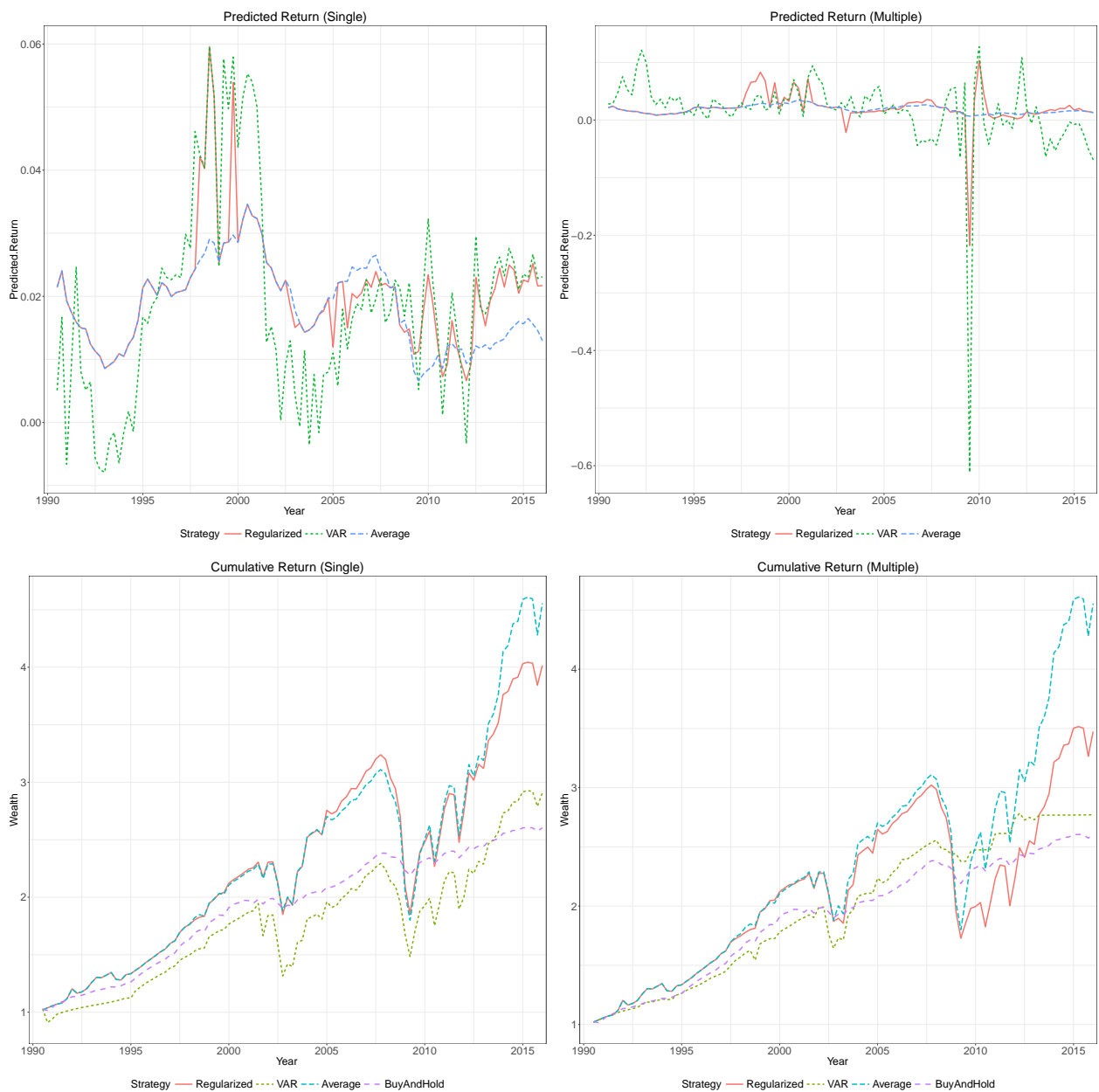}
\vspace{0.2in}

\small The regularization criterion is one-step ahead Predictive Cross-validation. The top two panels are the one-quarter ahead returns from both single and multiple predictor models. The least regularized prediction corresponds to the VAR prediction, and the most regularized prediction corresponds to the 80-quarter moving average. The bottom two panels are the cumulative returns for the quarterly updated mean-variance efficient portfolios by three strategies using both single and multiple predictor models. All procedures are updated using a rolling window of 80 quarters observations. The performance of the buy-and-hold portfolio is also plotted for comparison.
\end{figure}

The top two panels of Figure \ref{Fig3a} are the one-quarter-ahead returns from both single- and multiple-predictor models using predictive cross-validation. The least regularized prediction corresponds to the VAR prediction, and the most regularized prediction corresponds to the 80-quarter moving average. Though the range of return forecasts are mostly nonzero from 1990 to 2015, we find the forecasts are not robust to the regularization specification. We find one interesting pattern--that, in the multiple-predictor plot, the one-step-ahead optimal forecast has a high co-movement with the VAR forecast during the periods of the internet bubble and 2008 financial crisis. In other periods, the one-step-ahead optimal forecast almost overlaps with the moving average forecast.

The bottom two panels of Figure \ref{Fig3a} plot the wealth evolution of one dollar invested in a market-timing strategy using three procedures. The lines are the cumulative returns for the quarterly updated mean-variance efficient portfolios using both single- and multiple-predictor models. All procedures are updated using a rolling window of 80 quarters' observations. We also plot the performance of the buy-and-hold portfolio for comparison. We find substantial evidence of return predictability that all three strategies outperform the buy-and-hold portfolio. 

Table \ref{Tab2} lists the portfolio return distribution and finds the portfolio performances are sensitive to the regularization specification. The portfolio return distributions match the bias-variance intuition that the least regularized forecasts (VAR) tend to have a large range and possibly a smaller bias, whereas the most regularized forecast (moving average) has the lowest standard error. In the period from 1990 to 2015, the moving-average strategy has the best performance in this period, followed by the one-step-ahead optimal strategy. 

\begin{figure}
\caption{Market Timing Strategy (AIC Selection) \label{Fig3b}}
\vspace{0.2in}
\includegraphics[width=\textwidth]{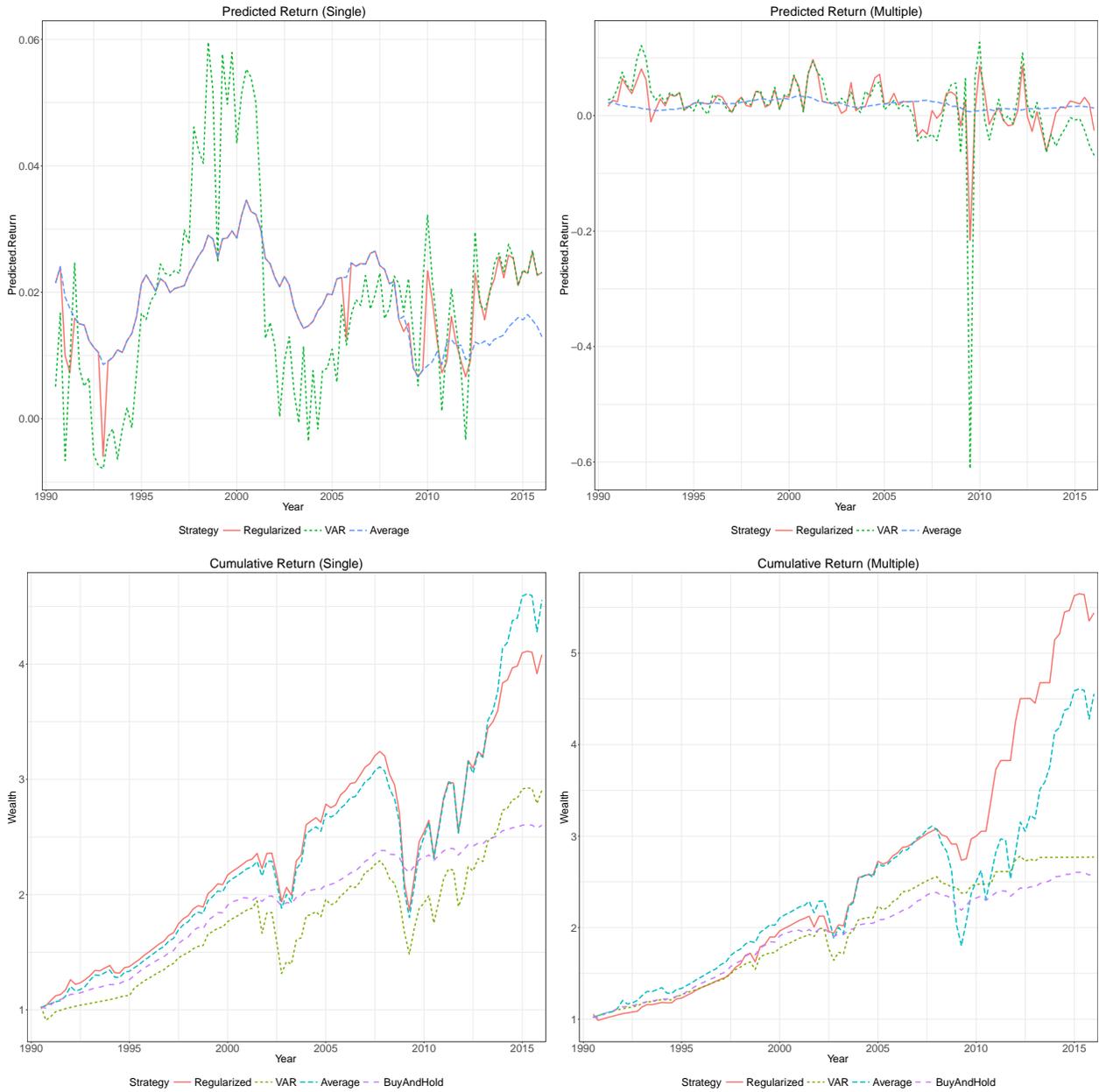}
\vspace{0.2in}

\small The regularization criterion is AIC selection. The top two panels are the one-quarter ahead returns from both single and multiple predictor models. The least regularized prediction corresponds to the VAR prediction, and the most regularized prediction corresponds to the 80-quarter moving average. The bottom two panels are the cumulative returns for the quarterly updated mean-variance efficient portfolios by three strategies using both single and multiple predictor models. All procedures are updated using a rolling window of 80 quarters observations. The performance of the buy-and-hold portfolio is also plotted for comparison.
\end{figure}

For a robustness check, we also repeat the same exercise in Figure \ref{Fig3b} using a modified AIC as the model-selection criterion, which is not necessarily optimal for the one-step-ahead forecast. The penalized model size is the number of predictors selected. We see similar time-series patterns of all three one-step-ahead forecasts as well as the non-sensitivity. We also see all three strategies outperform the fixed-split portfolio buy-and-hold strategy. However, we find a strong performance of the optimal strategy using multiple predictors. In the recent period after the 2008 financial crisis, this strategy's strong outperformance supports the evidence of using some predictors such as book-to-market ratio, CAY, and T-bill rates.

\subsection{Bond Premia Prediction \label{sec42}}

Since the failure of the expectations hypothesis, the literature focus on forecasting the variation in one-year expected excess returns for bonds of multiple maturities. \cite{ludvigson2009macro} build an empirical linkage between cyclical fluctuations in excess bond returns and macroeconomic fundamentals in a dynamic factor model. They use a small number of principal components instead of observed macroeconomic predictors in the predictive regressions and find these latent factors associated with real economic activity have significant predictive power beyond financial predictors. 

Our exercise investigates the sensitivity of their findings with special attention paid to the joint prediction sensitivity. The factors that explain most of the variation on the right-hand side need not be the same as the factors most important for predicting the left-hand side. For the same robustness check, \cite{ludvigson2010factor} estimate a dynamic factor model for each of the eight subgroups using a Bayesian procedure, which can also be quickly implemented using our procedure.

Let $rx_{t+1}$ denote the continuously compounded (log) excess return on an n-year discount bond in period t + 1. We study the log 1-year holding period returns for two- to five- bond over the log yield of one-year bond from January 1964 to December 2003. Below is the SUR  system for the joint prediction. For $rx_{t+1}^\intercal = \bmat{rx_{t+1,1}^\intercal, \cdots, rx_{t+1,4}^\intercal} $, 
\[ rx_{t+1} = X_tB + E_{t+1}, \]
where $X_{t} = \mbox{diag}\{X_{t,1}, \cdots, X_{t,4}\}$, $B^\intercal = \bmat{\beta_1^\intercal, \cdots, \beta_4^\intercal}$, and $E_{t+1} = \bmat{\epsilon_{t+1,1}^\intercal, \cdots, \epsilon_{t+1,4}^\intercal} \sim \mathcal{N}(0, \Sigma \otimes I)$.

The SUR implementation is the same as the VAR system and we stack equations to form a univariate linear model. The difference is, in addition to the standard Lasso and Ridge regularization on $B$, we can apply a group regularization in the perspective of \cite{yuan2006model}. This group regularization can be implemented as a multivariate gaussian model in the package [\texttt{glmnet}]. An attractive property is the group factor selection from the group Lasso regularization, then we can see the common macro predictors in excess returns for bonds of multiple maturities. 
\[ B^\star = \argmin_{B} l(B \mid \Sigma) + \lambda \sum^p_{k=1} \sum^4_{j=1}\left|\beta_{j,k}\right|. \]

\begin{figure}
\caption{Regularized Seemingly Unrelated Regressions Forecasts \label{Fig4}}
\vspace{0.2in}
\includegraphics[width=\textwidth]{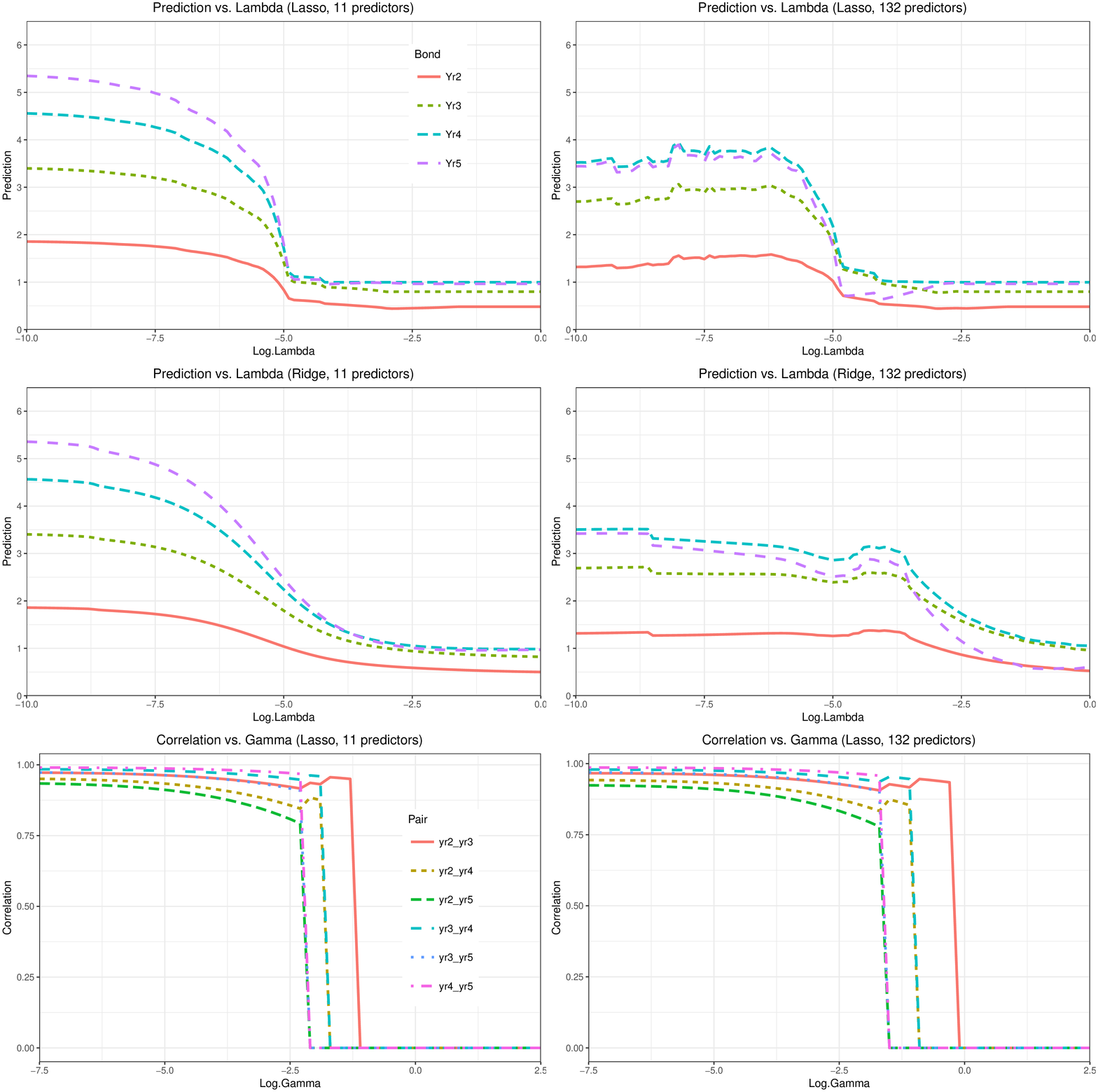}
\vspace{0.2in}

\small The figure reports the predicted value regularization paths for excess bond returns on Dec. 2003 in a SUR model. The left panel uses the CP factor and the 10 factors from PCA in \cite{ludvigson2009macro}, while the right panel uses the CP factor and the 131 macro fundamentals. Lasso and Ridge regularization results are provided. The shock correlation for the cross-section of excess bond returns are also provided. 
\end{figure}

\begin{figure}
\caption{Group Regularized SUR Forecasts \label{Fig5}}
\vspace{0.2in}
\includegraphics[width=\textwidth]{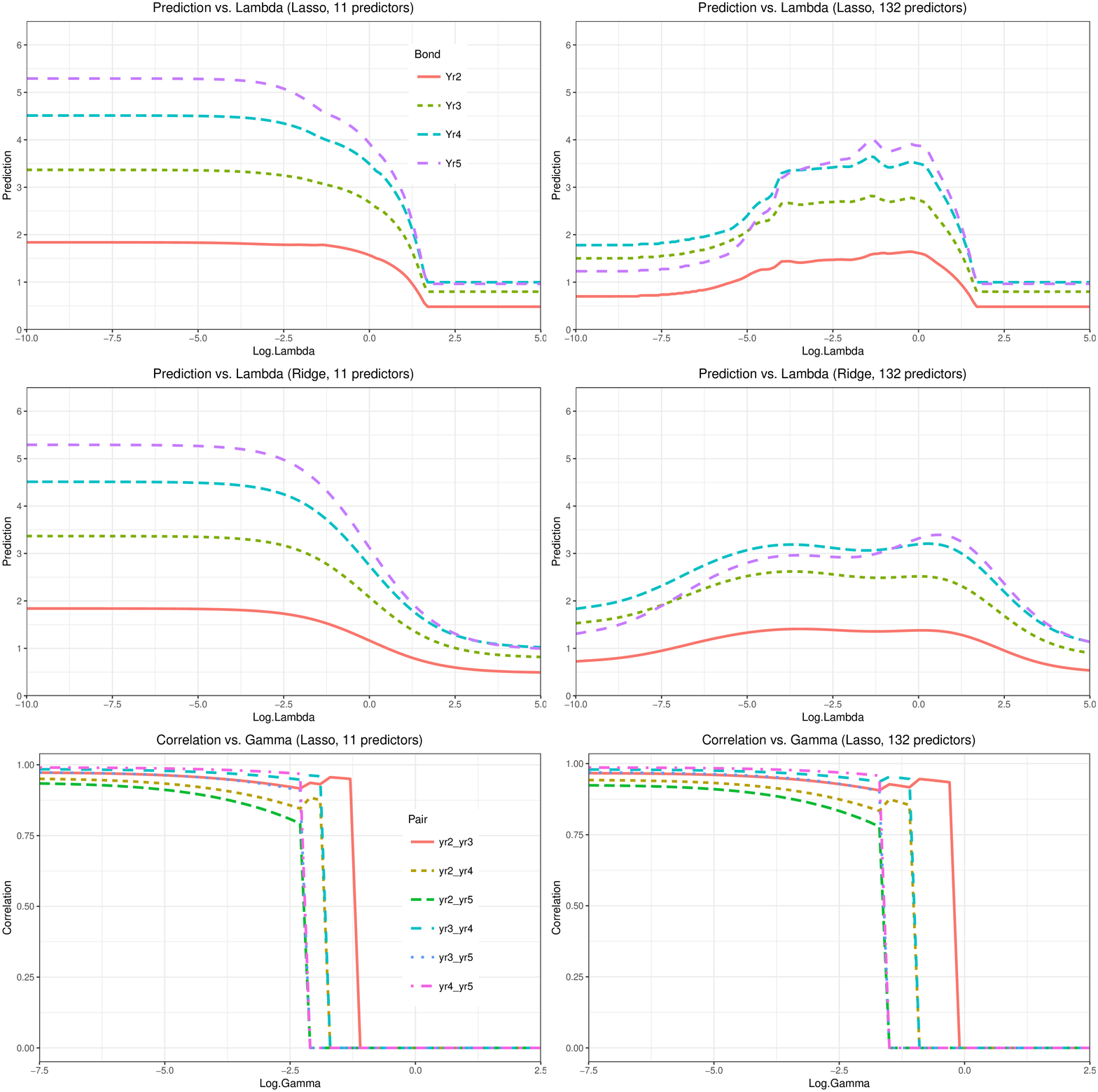}
\vspace{0.2in}

\small The figure reports the predicted value regularization paths for excess bond returns on Dec. 2003 in a group regularized SUR model. The left panel uses the CP factor and the 10 factors from PCA in \cite{ludvigson2009macro}, while the right panel uses the CP factor and the 131 macro fundamentals. Group Lasso and Ridge regularization results are provided. The shock correlation for the cross-section of excess bond returns are also provided. 
\end{figure}

\begin{figure}
\caption{Macro Factor Selection \label{Fig6}}
\vspace{0.2in}
\includegraphics[width=\textwidth, height=0.8\textheight]{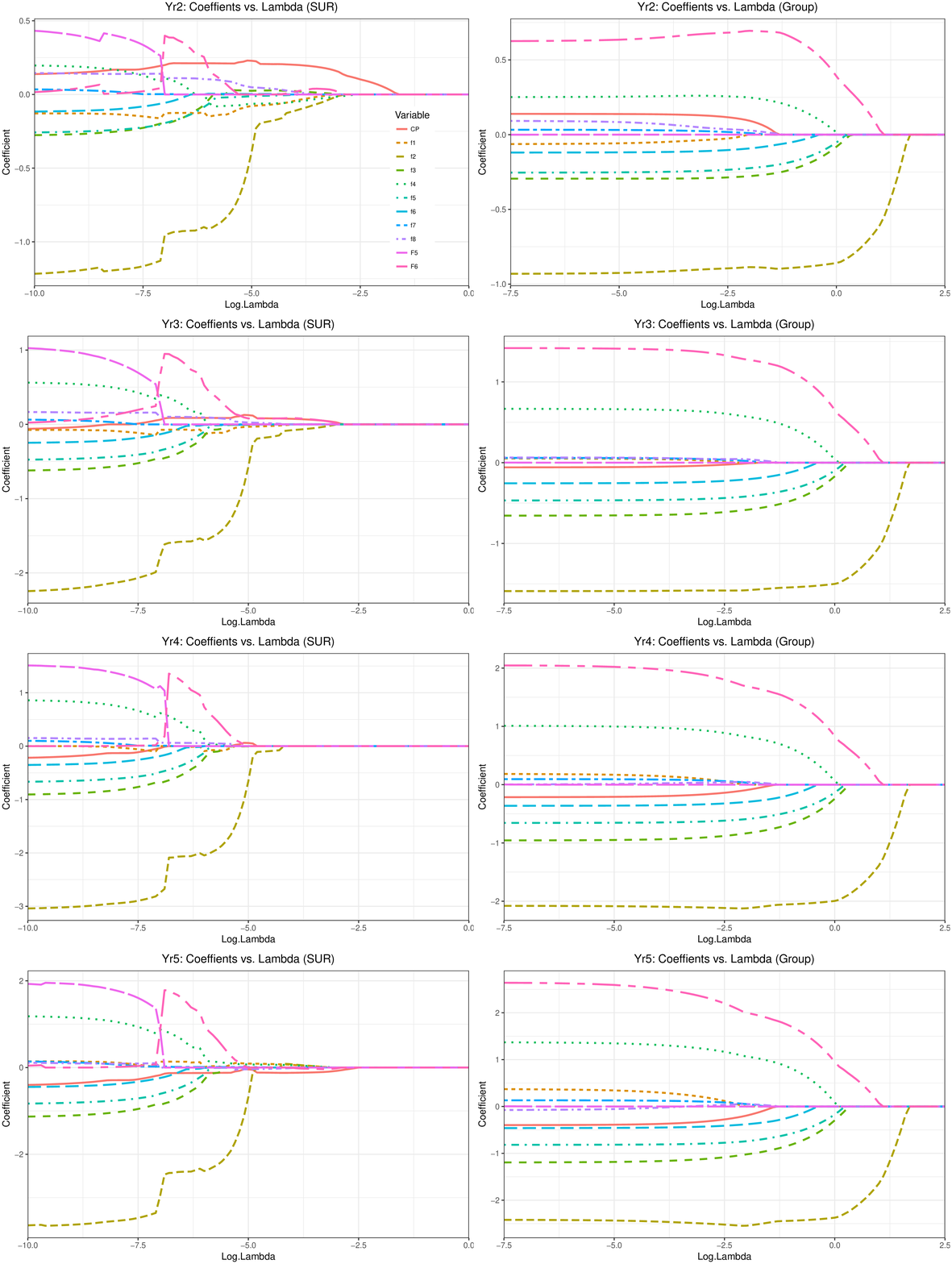}
\vspace{0.2in}

\small The figure reports the predictor coefficients regularization paths for excess bond returns on Dec. 2003 using one-month lag predictors. There are the CP factor and the 10 factors from PCA in \cite{ludvigson2009macro}. The left panel shows the SUR model and the right panel shows the group regularized SUR model. All plots share the same legend in the (1,1) subfigure. 
\end{figure}

We have two models for one-month ahead prediction comparison. The first model includes the Cochrane-Piazzesi factor, a linear combination of five forward spreads, plus eight principal components from \cite{ludvigson2009macro} as well as two single factors constructed as a linear combination of five and six estimated factors. The first model includes the Cochrane-Piazzesi factor and a panel of 131 monthly macroeconomic time series. We show the regularization path for SUR forecasts in Figure \ref{Fig4} and the group regularized SUR forecasts in Figure \ref{Fig5}. In Figure \ref{Fig6}, we show the regularization path for predictor loadings in the first 11-predictor model.

We can see the significant variation for predicted value for the Dec. 2003. In both Figure  \ref{Fig4} and \ref{Fig5}, we can see the sensitive change for prediction over a wide range of tuning parameters. Though the 11-predictor model has a smoother prediction over the 132-predictor model, the predicted values for the same month are entirely different for all four bonds in both SUR and group regularized SUR models. The latent factor extraction does not seem a robust approach concerning different priors, and useful macro information could be lost due to the ad-hoc decision in dimension reduction (or change of the shrinkage prior hyper-parameter). 

Furthermore, in the bottom two plots, we can see it is more efficient to study the prediction in an SUR framework because the correlations of the shocks among four bonds are high with a low penalty. After hundreds of predictors, the cross-sectional correlations among bonds are still high. With the Cochrane-Piazzesi factor and macroeconomic series, there are still considerable co-movement in the regularization path for predicted values and shock correlations. 

However, in the SUR prediction, the predictive factors have different strengths and existence for various bonds. The top three predictors stand out include f2 (2nd PC), CP (Cochrane-Piazzesi), and F6 (the single factor constructed as a linear combination of six estimated factors in \cite{ludvigson2009macro}). An attractive feature for the group regularized SUR prediction is the common factor selection in the cross-sectional of bond excess returns. The regularization path is similar to every bond with the top two predictors f2 and F6. Therefore, we can see the supervised learning is robust for a different penalty or prior distribution specification for bond prediction.

\section{Discussion \label{sec5}}
By exploiting the fundamental duality between regularization penalties and prior distributions, we provide a MAP approach that jointly regularizes both expected values and variance-covariance matrices for the high-dimensional VAR and SUR systems. Also, we use an iterative proximal algorithm that solves two convex subproblems on the predictor coefficients and the variance-covariance matrix for maximizing the posterior mode or penalized likelihood. Moreover, for the curse of dimensionality about the variance-covariance matrix, we introduce a Lasso-tilted inverse Wishart prior for regularization. 

Our regularization approach offers several computational and empirical advantages, which include MAP's computational convenience over traditional MCMC procedures, the regularization-path plots for prior sensitivity analysis to forecasting power, and the possibility of building a high-dimension VAR for the model uncertainty and feedback effects. Furthermore, a regularization penalty using shrinkage priors, such as double-exponential distribution, provides many new empirical insights of variable selection in the Bayesian predictive analysis. 

In the equity-premium prediction example, we demonstrate the change in out-of-sample prediction performance and orthogonal impulse response due to the change of specification in shrinkage priors. We find the risk premium forecasts are sensitive to the regularization penalty or the prior hyper-parameters in the Bayesian language. We also find significant predictability of excess S\&P 500 returns only using book-to-market ratio, CAY, and T-bill rates when implementing the market-timing strategies. 

The trade-off is we obtain only a posterior-mode prediction instead of the full posterior distribution. Consequently, one caveat is we do not fully account for parameter uncertainty as in \cite{Kandel1996} and \cite{Barberis2000}. Although our results are quantitatively similar to their full Bayesian analysis, our regularized estimation can be computationally faster than MCMC procedures. Moreover, we can easily model the parameter sparsity in our Lasso specification, and we provide full prior sensitivity analysis rather having to specify the prior hyper-parameters. 

Our approach builds on previously underexploited relationships between prior hyper-parameter selection and out-of-sample prediction power. In empirical finance and macroeconomics, most predictors lack statistical and even economic significance. For example, \cite{feng2017taming} provide a post-selection inference method to tame the zoo of factors in the cross-sectional asset pricing literature. For these problems involving large variable selection, our regularization procedure provides a lens to link the model uncertainty and prediction power. One possible direction is to apply the SUR system in the cross-sectional returns to extract common factors. 

When implementing the elastic net package for variable selection, statisticians usually standardized the variables into the same scale. Otherwise, predictors with higher variance have smaller coefficients and therefore less penalization. It is equivalent to set a vector of heterogeneous hyper-parameter $\{\lambda_p\}$ of the ``informative" prior distribution for every predictor p. In our empirical examples, we follow the economic literature instead and do not standardize the predictors to have a direct comparison with the current empirical findings. Researchers use our regularization method should be aware of this variable selection issue.  

In the bond prediction study, we find a significant information loss for the ad-hoc decision in dimension reduction technique. We find the predictions from their latent factor models to be sensitive to prior specifications. However, the shrinkage prior for predictor selection is robust for the penalty or prior distribution specification. The SUR system is another underexploited model in the empirical finance literature to explore the cross-sectional signals. There are some directions possible for future research, including the time-varying specification for the existence and strength of common predictors in the Bayesian regularization framework. 

\bibliography{bayesReg.bib}

\begin{thebibliography}{}

\bibitem[\protect\citeauthoryear{Ba{\'n}bura, Giannone, and
  Reichlin}{Ba{\'n}bura et~al.}{2010}]{banbura2010large}
Ba{\'n}bura, M., D.~Giannone, and L.~Reichlin (2010).
\newblock Large {B}ayesian vector auto regressions.
\newblock {\em Journal of Applied Econometrics\/}~{\em 25\/}(1), 71--92.

\bibitem[\protect\citeauthoryear{Barberis}{Barberis}{2000}]{Barberis2000}
Barberis, N. (2000).
\newblock Investing for the long run when returns are predictable.
\newblock {\em The Journal of Finance\/}~{\em 55\/}(1), 225--264.

\bibitem[\protect\citeauthoryear{Bien and Tibshirani}{Bien and
  Tibshirani}{2011}]{Bien2011}
Bien, J. and R.~J. Tibshirani (2011).
\newblock Sparse estimation of a covariance matrix.
\newblock {\em Biometrika\/}~{\em 98\/}(4), 807.

\bibitem[\protect\citeauthoryear{Cai and Liu}{Cai and
  Liu}{2011}]{cai2011adaptive}
Cai, T. and W.~Liu (2011).
\newblock Adaptive thresholding for sparse covariance matrix estimation.
\newblock {\em Journal of the American Statistical Association\/}~{\em
  106\/}(494), 672--684.

\bibitem[\protect\citeauthoryear{Campbell and Shiller}{Campbell and
  Shiller}{1988}]{Campbell1989}
Campbell, J.~Y. and R.~J. Shiller (1988).
\newblock The dividend-price ratio and expectations of future dividends and
  discount factors.
\newblock {\em {R}eview of {F}inancial {S}tudies\/}~{\em 1\/}(3), 195--228.

\bibitem[\protect\citeauthoryear{Campbell and Shiller}{Campbell and
  Shiller}{1991}]{campbell1991yield}
Campbell, J.~Y. and R.~J. Shiller (1991).
\newblock Yield spreads and interest rate movements: A bird's eye view.
\newblock {\em The Review of Economic Studies\/}~{\em 58\/}(3), 495--514.

\bibitem[\protect\citeauthoryear{Campbell and Thompson}{Campbell and
  Thompson}{2008}]{Campbell2008}
Campbell, J.~Y. and S.~B. Thompson (2008).
\newblock Predicting excess stock returns out of sample: Can anything beat the
  historical average?
\newblock {\em {R}eview of {F}inancial {S}tudies\/}~{\em 21\/}(4), 1509--1531.

\bibitem[\protect\citeauthoryear{Carriero, Kapetanios, and Marcellino}{Carriero
  et~al.}{2012}]{carriero2012forecasting}
Carriero, A., G.~Kapetanios, and M.~Marcellino (2012).
\newblock Forecasting government bond yields with large {B}ayesian vector
  autoregressions.
\newblock {\em Journal of Banking \& Finance\/}~{\em 36\/}(7), 2026--2047.

\bibitem[\protect\citeauthoryear{Chan, Eisenstat, and Koop}{Chan
  et~al.}{2016}]{chan2016large}
Chan, J.~C., E.~Eisenstat, and G.~Koop (2016).
\newblock Large {B}ayesian varmas.
\newblock {\em Journal of Econometrics\/}~{\em 192\/}(2), 374--390.

\bibitem[\protect\citeauthoryear{Chen and Huang}{Chen and
  Huang}{2012}]{chen2012sparse}
Chen, L. and J.~Z. Huang (2012).
\newblock Sparse reduced-rank regression for simultaneous dimension reduction
  and variable selection.
\newblock {\em Journal of the American Statistical Association\/}~{\em
  107\/}(500), 1533--1545.

\bibitem[\protect\citeauthoryear{Cochrane}{Cochrane}{2008}]{Cochrane2008}
Cochrane, J.~H. (2008).
\newblock The dog that did not bark: A defense of return predictability.
\newblock {\em {R}eview of {F}inancial {S}tudies\/}~{\em 21\/}(4), 1533--1575.

\bibitem[\protect\citeauthoryear{Cochrane and Piazzesi}{Cochrane and
  Piazzesi}{2005}]{cochrane2005bond}
Cochrane, J.~H. and M.~Piazzesi (2005).
\newblock Bond risk premia.
\newblock {\em American Economic Review\/}~{\em 95\/}(1), 138--160.

\bibitem[\protect\citeauthoryear{Das and Dey}{Das and Dey}{2010}]{Das2010}
Das, S. and D.~K. Dey (2010).
\newblock On {B}ayesian inference for generalized multivariate gamma
  distribution.
\newblock {\em Statistics \& Probability Letters\/}~{\em 80\/}(19), 1492--1499.

\bibitem[\protect\citeauthoryear{Davis, Zang, and Zheng}{Davis
  et~al.}{2016}]{Davis2015}
Davis, R.~A., P.~Zang, and T.~Zheng (2016).
\newblock Sparse vector autoregressive modeling.
\newblock {\em Journal of Computational and Graphical Statistics\/}~{\em
  25\/}(4), 1077--1096.

\bibitem[\protect\citeauthoryear{Fama and Bliss}{Fama and
  Bliss}{1987}]{fama1987information}
Fama, E.~F. and R.~R. Bliss (1987).
\newblock The information in long-maturity forward rates.
\newblock {\em The American Economic Review\/}, 680--692.

\bibitem[\protect\citeauthoryear{Fama and French}{Fama and
  French}{1988}]{Fama1988}
Fama, E.~F. and K.~R. French (1988).
\newblock Dividend yields and expected stock returns.
\newblock {\em {J}ournal of {F}inancial {E}conomics\/}~{\em 22\/}(1), 3--25.

\bibitem[\protect\citeauthoryear{Fan, Liao, and Mincheva}{Fan
  et~al.}{2013}]{fan2013large}
Fan, J., Y.~Liao, and M.~Mincheva (2013).
\newblock Large covariance estimation by thresholding principal orthogonal
  complements.
\newblock {\em Journal of the Royal Statistical Society: Series B\/}~{\em
  75\/}(4), 603--680.

\bibitem[\protect\citeauthoryear{Feng, Giglio, and Xiu}{Feng
  et~al.}{2017}]{feng2017taming}
Feng, G., S.~Giglio, and D.~Xiu (2017).
\newblock Taming the factor zoo.
\newblock Technical report, City University of Hong Kong.

\bibitem[\protect\citeauthoryear{George and Foster}{George and
  Foster}{2000}]{george2000calibration}
George, E. and D.~P. Foster (2000).
\newblock Calibration and empirical bayes variable selection.
\newblock {\em Biometrika\/}~{\em 87\/}(4), 731--747.

\bibitem[\protect\citeauthoryear{Giannone, Lenza, and Primiceri}{Giannone
  et~al.}{2015}]{giannone2015prior}
Giannone, D., M.~Lenza, and G.~E. Primiceri (2015).
\newblock Prior selection for vector autoregressions.
\newblock {\em {R}eview of {E}conomics and {S}tatistics\/}~{\em 97\/}(2),
  436--451.

\bibitem[\protect\citeauthoryear{Ishwaran and Rao}{Ishwaran and
  Rao}{2005}]{ishwaran2005spike}
Ishwaran, H. and J.~S. Rao (2005).
\newblock Spike and slab variable selection: frequentist and {B}ayesian
  strategies.
\newblock {\em Annals of Statistics\/}~{\em 33\/}(2), 730--773.

\bibitem[\protect\citeauthoryear{Jurado, Ludvigson, and Ng}{Jurado
  et~al.}{2015}]{jurado2015measuring}
Jurado, K., S.~C. Ludvigson, and S.~Ng (2015).
\newblock Measuring uncertainty.
\newblock {\em The American Economic Review\/}~{\em 105\/}(3), 1177--1216.

\bibitem[\protect\citeauthoryear{Kandel and Stambaugh}{Kandel and
  Stambaugh}{1996}]{Kandel1996}
Kandel, S. and R.~F. Stambaugh (1996).
\newblock On the predictability of stock returns: an asset-allocation
  perspective.
\newblock {\em The Journal of Finance\/}~{\em 51\/}(2), 385--424.

\bibitem[\protect\citeauthoryear{Ludvigson and Ng}{Ludvigson and
  Ng}{2010}]{ludvigson2010factor}
Ludvigson, S. and S.~Ng (2010).
\newblock A factor analysis of bond risk premia.
\newblock In {\em Handbook of empirical economics and finance}, Volume~1, pp.\
  313 -- 372.

\bibitem[\protect\citeauthoryear{Ludvigson and Ng}{Ludvigson and
  Ng}{2009}]{ludvigson2009macro}
Ludvigson, S.~C. and S.~Ng (2009).
\newblock Macro factors in bond risk premia.
\newblock {\em The Review of Financial Studies\/}~{\em 22\/}(12), 5027--5067.

\bibitem[\protect\citeauthoryear{Merton}{Merton}{1969}]{Merton1969}
Merton, R.~C. (1969).
\newblock Lifetime portfolio selection under uncertainty: The continuous-time
  case.
\newblock {\em The {R}eview of {E}conomics and {S}tatistics\/}~{\em 51\/}(3),
  247--257.

\bibitem[\protect\citeauthoryear{Nicholson, Matteson, and Bien}{Nicholson
  et~al.}{2015}]{Nicholson2015}
Nicholson, W., D.~Matteson, and J.~Bien (2015).
\newblock Varx-l: Structured regularization for large vector autoregressions
  with exogenous variables.
\newblock Technical report, Cornell University.

\bibitem[\protect\citeauthoryear{Park and Casella}{Park and
  Casella}{2008}]{Park2008}
Park, T. and G.~Casella (2008).
\newblock The {B}ayesian lasso.
\newblock {\em Journal of the American Statistical Association\/}~{\em
  103\/}(482), 681--686.

\bibitem[\protect\citeauthoryear{Polson and Scott}{Polson and
  Scott}{2015}]{polson2015mixtures}
Polson, N.~G. and J.~G. Scott (2015).
\newblock Mixtures, envelopes and hierarchical duality.
\newblock {\em Journal of the Royal Statistical Society: Series B\/}~{\em
  78\/}(4), 701--727.

\bibitem[\protect\citeauthoryear{Polson, Scott, and Willard}{Polson
  et~al.}{2015}]{polson2015proximal}
Polson, N.~G., J.~G. Scott, and B.~T. Willard (2015).
\newblock Proximal algorithms in statistics and machine learning.
\newblock {\em Statistical Science\/}~{\em 30\/}(4), 559--581.

\bibitem[\protect\citeauthoryear{Rapach, Strauss, and Zhou}{Rapach
  et~al.}{2010}]{Rapach2010}
Rapach, D.~E., J.~K. Strauss, and G.~Zhou (2010).
\newblock Out-of-sample equity premium prediction: Combination forecasts and
  links to the real economy.
\newblock {\em {R}eview of {F}inancial {S}tudies\/}~{\em 23\/}(2), 821--862.

\bibitem[\protect\citeauthoryear{Rothman, Levina, and Zhu}{Rothman
  et~al.}{2010}]{rothman2010sparse}
Rothman, A.~J., E.~Levina, and J.~Zhu (2010).
\newblock Sparse multivariate regression with covariance estimation.
\newblock {\em Journal of Computational and Graphical Statistics\/}~{\em
  19\/}(4), 947--962.

\bibitem[\protect\citeauthoryear{Samuelson}{Samuelson}{1969}]{Samuelson1969}
Samuelson, P.~A. (1969).
\newblock Lifetime portfolio selection by dynamic stochastic programming.
\newblock {\em The {R}eview of {E}conomics and {S}tatistics\/}~{\em 51\/}(3),
  239--246.

\bibitem[\protect\citeauthoryear{Song and Bickel}{Song and
  Bickel}{2011}]{Song2011}
Song, S. and P.~J. Bickel (2011).
\newblock Large vector auto regressions.
\newblock Technical report, University of California, Berkeley.

\bibitem[\protect\citeauthoryear{Stock and Watson}{Stock and
  Watson}{2002a}]{stock2002forecasting}
Stock, J.~H. and M.~W. Watson (2002a).
\newblock Forecasting using principal components from a large number of
  predictors.
\newblock {\em Journal of the American statistical association\/}~{\em
  97\/}(460), 1167--1179.

\bibitem[\protect\citeauthoryear{Stock and Watson}{Stock and
  Watson}{2002b}]{Stock2002}
Stock, J.~H. and M.~W. Watson (2002b).
\newblock Macroeconomic forecasting using diffusion indexes.
\newblock {\em Journal of Business \& Economic Statistics\/}~{\em 20\/}(2),
  147--162.

\bibitem[\protect\citeauthoryear{Welch and Goyal}{Welch and
  Goyal}{2008}]{Welch2008}
Welch, I. and A.~Goyal (2008).
\newblock A comprehensive look at the empirical performance of equity premium
  prediction.
\newblock {\em {R}eview of {F}inancial {S}tudies\/}~{\em 21\/}(4), 1455--1508.

\bibitem[\protect\citeauthoryear{Xu and Ghosh}{Xu and
  Ghosh}{2015}]{xu2015bayesian}
Xu, X. and M.~Ghosh (2015).
\newblock Bayesian variable selection and estimation for group lasso.
\newblock {\em Bayesian Analysis\/}~{\em 10\/}(4), 909--936.

\bibitem[\protect\citeauthoryear{Yuan and Lin}{Yuan and
  Lin}{2005}]{yuan2005efficient}
Yuan, M. and Y.~Lin (2005).
\newblock Efficient empirical bayes variable selection and estimation in linear
  models.
\newblock {\em Journal of the American Statistical Association\/}~{\em
  100\/}(472), 1215--1225.

\bibitem[\protect\citeauthoryear{Yuan and Lin}{Yuan and
  Lin}{2006}]{yuan2006model}
Yuan, M. and Y.~Lin (2006).
\newblock Model selection and estimation in regression with grouped variables.
\newblock {\em Journal of the Royal Statistical Society: Series B\/}~{\em
  68\/}(1), 49--67.

\bibitem[\protect\citeauthoryear{Zantedeschi, Damien, and Polson}{Zantedeschi
  et~al.}{2011}]{Zantedeschi2011}
Zantedeschi, D., P.~Damien, and N.~G. Polson (2011).
\newblock Predictive macro-finance with dynamic partition models.
\newblock {\em Journal of the American Statistical Association\/}~{\em
  106\/}(494), 427--439.

\bibitem[\protect\citeauthoryear{Zellner}{Zellner}{1962}]{Zellner1962}
Zellner, A. (1962).
\newblock An efficient method of estimating seemingly unrelated regressions and
  tests for aggregation bias.
\newblock {\em Journal of the American statistical Association\/}~{\em
  57\/}(298), 348--368.

\bibitem[\protect\citeauthoryear{Zellner}{Zellner}{1963}]{zellner1963estimators}
Zellner, A. (1963).
\newblock Estimators for seemingly unrelated regression equations: Some exact
  finite sample results.
\newblock {\em Journal of the American Statistical Association\/}~{\em
  58\/}(304), 977--992.

\bibitem[\protect\citeauthoryear{Zellner}{Zellner}{1971}]{Zellner1971}
Zellner, A. (1971).
\newblock {\em An Introduction to Bayesian Inference in Econometrics}.
\newblock Wiley, New York.

\bibitem[\protect\citeauthoryear{Zellner and Ando}{Zellner and
  Ando}{2010}]{Zellner2010}
Zellner, A. and T.~Ando (2010).
\newblock A direct monte carlo approach for {B}ayesian analysis of the
  seemingly unrelated regression model.
\newblock {\em Journal of Econometrics\/}~{\em 159\/}(1), 33--45.

\end{thebibliography}

\newpage

\begin{table}[]
\begin{center}
\caption{Predictor Description. \label{Tab0}}
\vspace{0.2in}
\resizebox{\textwidth}{!}{%
\begin{tabular}{@{}ll@{}}
\toprule
Predictor & Description \\ \midrule
Dividend Yield & Difference between the log of dividends and the log of lagged prices. \\
Earning Price Ratio & Difference between the log of earnings and the log of prices. \\
Book to Market Ratio & Ratio of book value to market value for the Dow Jones Industrial Average. \\
Dividend Payout Ratio & Difference between the log of dividends and the log of earnings. \\
T-bill rates & 3- Month Treasury Bill \\
Long Term Rate of Return & Long term yield on government bonds. \\
Default Return Spread & Difference between long-term corporate and government bond returns. \\
Investment to Capital Ratio & Ratio of aggregate investment to aggregate capital for the whole economy. \\
CAY & Consumption, wealth, income ratio \\ \bottomrule
\end{tabular}
}
	
\end{center}
\vspace{0.2in}

\small 
The predictor variables used in Section \ref{sec41} are defined above. Full details of variable definitions and sources are given in \cite{Welch2008}. Our quarterly data is from 1952 to 2015.

\end{table}

\begin{table}
\begin{center}
\caption{Comparison between Bayesian Analysis and Regularization.\label{Tab1}}
\vspace{0.2in}
\begin{tabular}{cccccc}
\hline
 & \multicolumn{2}{c}{Barberis} &  & \multicolumn{2}{c}{Regularization} \\ \hline
\multirow{6}{*}{1952-2015} & a & B &  & \multicolumn{2}{c}{$\beta$} \\ \cline{2-3}\cline{5-6}
 & 6.764e-02 & 1.656e-2 &  & 6.498e-02 & 1.263e-02 \\
 & -6.167e-14 & 1.000 &  & -5.590e-02 & 9.591e-01 \\ \cline{2-3}\cline{5-6}
 & \multicolumn{2}{c}{$\Sigma$} &  & \multicolumn{2}{c}{$\Sigma$} \\ \cline{2-3}\cline{5-6}
 & 6.191e-03 & -1.700e-18 &  & 6.191e-03 & 1.173e-04 \\
 &  & 1.019e-27 &  &  & 6.158e-03 \\ \hline
\multirow{6}{*}{1986-2015} & a & B &  & \multicolumn{2}{c}{$\beta$} \\ \cline{2-3}\cline{5-6}
 & 1.417e-01 & 3.357e-02 &  & 4.145e-03 & 2.857e-02 \\
 & 1.293e-13  & 1.000 &  & 1.089e-02 & 9.1567e-01 \\ \cline{2-3}\cline{5-6}
 & \multicolumn{2}{c}{$\Sigma$} &  & \multicolumn{2}{c}{$\Sigma$} \\ \cline{2-3}\cline{5-6}
 & 6.362e-03 & -2.197e-18 &  & 6.397e-03 & 3.002e-05 \\
 &  & 5.110e-28 &  &  & 6.524e-03 \\ \cline{2-6} 
\end{tabular}
\end{center}
\vspace{0.2in}

\small This table shows the posterior estimates from the fully Bayesian analysis from \cite{Barberis2000} and our corresponding Ridge regularization MAP estimates. The first comparison is between B and $\beta_{[:, 2]}$, coefficients of D/P$_{t-1}$ on ret$_t$ and D/P$_{t}$. The second comparison is between the variance-covariance matrix. 
\end{table}

\begin{table}
\begin{center}
\caption{Comparison for Model Prediction Distribution \label{Tab2}}
\vspace{0.2in}
\begin{tabular}{@{}cccccccc@{}}
\toprule
             & CV.S & AIC.S & VAR.S & CV.M & AIC.M & VAR.M & Average \\ \midrule
Min.         & 0.05\%    & -1.40\%    & -2.35\%    & -21.84\%    & -21.65\%     & -61.19\%     & 0.05\%  \\
1st Qu.      & 0.78\%    & 0.75\%     & 0.21\%     & 0.60\%      & 0.20\%       & -0.65\%      & 0.80\%  \\
Median       & 1.21\%    & 1.32\%     & 1.06\%     & 1.13\%      & 1.26\%       & 1.19\%       & 1.24\%  \\
Mean         & 1.35\%    & 1.24\%     & 1.16\%     & 1.33\%      & 1.27\%       & 0.76\%       & 1.14\%  \\
3rd Qu.      & 1.80\%    & 1.76\%     & 2.23\%     & 1.82\%      & 2.81\%       & 3.10\%       & 1.49\%  \\
Max.         & 4.71\%    & 2.63\%     & 4.71\%     & 10.35\%     & 8.70\%       & 12.72\%      & 2.04\%  \\
Sd.          & 0.85\%    & 0.74\%     & 1.53\%     & 2.86\%      & 3.54\%       & 7.22\%       & 0.50\%  \\
Sharpe Ratio & 1.58      & 1.67       & 0.76       & 0.47        & 0.36         & 0.11         & 2.26    \\ \bottomrule
\end{tabular}
\end{center}
\vspace{0.2in}

\small This table shows the empirical distribution of model predictions for quarterly risk premium from 1990 to 2015 used in the market-timing analysis. All strategies (single or multiple predictors) are updated using data from a rolling window of 80 quarters. The single predictor approach uses dividend-price ratio. The regularization criteria include one-step ahead Predictive Cross-Validation and AIC selection. VAR is the regular VAR(1) estimation, and the average is the 80-quarter moving average.
\end{table}


\end{document}